\documentclass[journal]{IEEEtran}
\usepackage{cite}
\usepackage[pdftex]{graphicx}
\usepackage{amssymb,amsmath}
\usepackage{float}
\usepackage{bm}
\usepackage{mathtools}
\usepackage{amsmath}
\usepackage{array}
\usepackage{multirow}
\usepackage{stfloats}
\usepackage[utf8]{inputenc}
\usepackage[english]{babel}
\usepackage{array}
\usepackage{mathrsfs}
\usepackage{makecell}
\usepackage{tabularx,booktabs}

\usepackage[most]{tcolorbox}
\usepackage{color, colortbl}

\usepackage{algorithm,algpseudocode}
\makeatletter
\renewcommand{\Function}[2]{%
  \csname ALG@cmd@\ALG@L @Function\endcsname{#1}{#2}%
  \def\jayden@currentfunction{#1}%
}
\newcommand{\funclabel}[1]{%
  \@bsphack
  \protected@write\@auxout{}{%
    \string\newlabel{#1}{{\jayden@currentfunction}{\thepage}}%
  }%
  \@esphack
}
\makeatother

\begin{document}
\title{Modular Medium Voltage AC to Low Voltage DC Converter for Extreme Fast Charging Applications}

\author{M A~Awal, Iqbal~Husain, Md~Rashed~Hassan~Bipu, Oscar~Andr\'{e}s~Montes, Fei~Teng, \\Hao~Feng, Mehnaz~Khan, and Srdjan~Lukic 

FREEDM Systems Center, North Carolina State University, Raleigh, NC 27695, USA
\thanks{This work is supported by the US Department of Energy under award number DE-EE0008450 for the project titled ``Intelligent, Grid-Friendly 1~MVA Medium Voltage Extreme Fast Charger."}

\thanks{The authors are with the FREEDM Systems Center, North Carolina State University, Raleigh, NC 27695 USA (e-mail:, mawal@ncsu.edu; ihusain2@ncsu.edu; mhassan3@ncsu.edu; omontes2@ncsu.edu; fteng@ncsu.edu; hfeng6@ncsu.edu; makhan4@ncsu.edu; smlukic@ncsu.edu;).}
}

\markboth{Awal et. al.: Modular Medium Voltage AC to Low Voltage DC Converter for Extreme Fast Charging Applications}
{}

\maketitle

\begin{abstract}
  A modular and scalable converter for medium voltage (MV) AC to low voltage (LV) DC power conversion is proposed; single-phase-modules (SPMs), each consisting of an active-front-end (AFE) stage and an isolated DC-DC stage, are connected in input-series-output-parallel (ISOP) configuration to reach desired voltage and power capacity. In prior art, high-speed bidirectional communication among modules and a centralized controller is required to ensure module-level voltage and power balancing, which severely limits the scalability and practical realization of higher voltage and higher power systems. Moreover, large capacitors are used to suppress double-line-frequency voltage variations on the common MV DC bus shared by the AFE and the DC-DC stage originating from AC power pulsations through the SPMs. We propose a comprehensive controller which achieves voltage and power balancing using complete decentralized control of the DC-DC stages based on only local sensor feedback and the AFE stages are controlled using feedback of only the LV DC output. Furthermore, reduced capacitor requirement on the MV DC bus is achieved through design and control. The proposed method is validated through simulation and experimental results.               
\end{abstract}

\vspace{10pt}


\IEEEpeerreviewmaketitle

\bstctlcite{IEEEexample:BSTcontrol}

\section{Introduction}
\label{sec:introduction}
Extreme fast charging (XFC) is one of the key enabling technologies to reduce range anxiety associated with electrified vehicles. Commercial and near commercial single port DC ultra fast chargers with power capacities up to $350\text{kW}$ have been reported which connect to $380\text{V}$-$480\text{V}$ AC input, typically generated from a medium voltage AC (MVAC) distribution feeder using a dedicated line-frequency service transformer. Installation of such high power charging systems requires substantial electrical and infrastructure service upgrades such as the service transformer, ground surface condition, electrical wiring and conduits, permits, and administration. Consequently, constructing charging stations with multiple charging ports, such as Tesla supercharger station in Mountainview, California, rather than single ports makes more economic sense since the site construction overheads can be distributed over multiple ports. However, capability to connect directly to MVAC input, such as $4.2\text{kV}$ or $13.2\text{kV}$ can be achieved using a power electronics based solid-state-transformer (SST)\cite{sst1,sst2,sst3,sst4,sst5,sst6,sst7,sst8,sst9}. Historically, SSTs were proposed to replace line-frequency transformers for AC-to-AC conversion and are typically realized using a three-stage topology, i.e., an AC-DC stage, an isolated DC-DC stage, and a DC-AC stage \cite{sstDrB}. Such a three-stage configuration facilitates DC connectivity enabling direct integration of battery energy storage systems(BESS) and/or PV resources\cite{sstRev,sst3}. 

\begin{figure}[htb]
	\makebox[\linewidth][c]{\includegraphics[angle = 0, clip, trim=0cm 0cm 0cm 0cm,  width=0.5\textwidth]{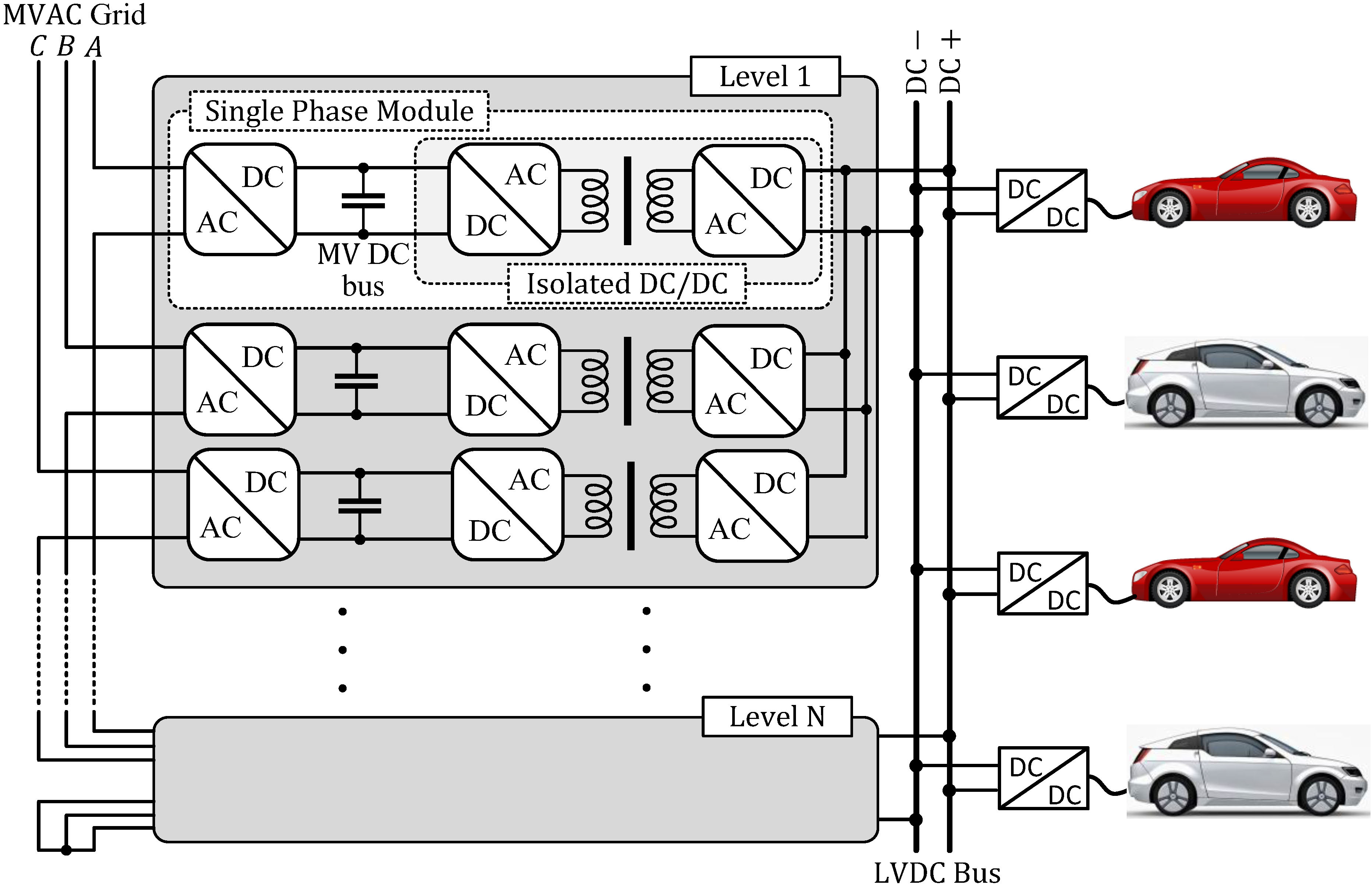}}
	\caption{An XFC station using a modular SST for delivering power from an MVAC feeder to an LVDC bus enabling multiple ultra-fast DC charging ports.}
	\label{fig:xfcStation}
\end{figure}

A modular SST configuration, excluding the DC-AC stage, can be utilized for construction of XFC stations, such as shown in Fig.~\ref{fig:xfcStation}, to avail direct connectivity to MVAC feeders. The input AC voltage for each single-phase module (SPM) is limited by the adopted power semiconductor device technology; series connection of N SPMs can be used in each phase to reach the desired input voltage. The outputs of the isolated DC-DC stages are tied in parallel to reach the desired power capacity.
Such a configuration offers modular and uniform construction of the power stage. However, scalability in terms of input voltage and overall charging capacity requires control and co-ordination among an increasing number of modules. Uniform power sharing among SPMs and equalizing/balancing the internal medium voltage DC (MVDC) buses during operation are the key challenges. In existing control approaches reported in literature, the $3N$ internal medium voltage DC (MVDC) buses are maintained by dynamically regulating the input AC current, whereas the low voltage DC (LVDC) bus is maintained by regulating the power flow through the $3N$ isolated DC-DC stages. Although intuitive, this approach presents two key challenges; first, for balanced three-phase operation one controllable input, such as the grid current, is used to regulate $3N$ MVDC buses. Second, $3N$ controllable inputs, such as the power flow through the DC-DC stages, are used to regulate a single output variable, i.e, the LVDC bus voltage. Consequently, voltage balancing among the internal MVDC buses and power flow balancing among the DC-DC stages become essential. A significant amount of research has been reported on the modules level voltage and power balancing \cite{balComp, cmplxMod, volBal1, powBal1, powBal2, sstDrB, sst9, balConHIerarchical}. In\cite{balComp}, the authors theoretically explain the need for explicit balancing methods and subsequently present a comparison between balancing through the active front end (AFE) stages and that through the isolated DC-DC stages. In \cite{sstDrB}, a centralized controller determines a common modulating duty ratio for the AFE stages based on the average of MV DC bus voltage measurements of all SPMs; the common duty ratio is locally adjusted by each SPM to compensate any bus voltage drift from the average value. The DC-DC stages are operated to follow a common power reference generated by the centralized controller and subsequently local compensations are performed to adjust any drift from the average power flow. In \cite{sst9}, the DC-DC stages use decentralized control but the MVDC bus voltage balancing is performed by a central controller. In\cite{balConHIerarchical}, the authors designate one of the DC-DC stages as the master and others are operated as slaves following the modulation signal generated by the master, whereas the AFE bridges are controlled based on the average MVDC bus voltage; to ensure voltage and power flow balance among modules a two step compensation strategy involving a centralized controller and module-level controllers is adopted. 
Overall, the existing methods employ complex balancing algorithms leveraging bidirectional high-speed communication among a centralized controller and the modules. Dependence on high-speed communication for real-time control is one of the major limiting factors for scalability. Evidently, there is a clear need for simplified control architecture with reduced communication requirement. Moreover, the MVDC bus voltages are subject to a double-line-frequency pulsation due to the AC and DC power flow through the AFE and the DC-DC stages of the SPMs, respectively. Large capacitors are used on the MVDC bus as energy buffers to suppress such voltage pulsations. Large capacitors rated for MV operation adds significant cost. Furthermore, relevant safety standards mandate fast discharge/bleeding of stored energy under fault condition and/or for maintenance \cite{UL}. Consequently, large energy storage capacitors on the internal MVDC buses complicates system level design. 

In this work, we propose a comprehensive design and implementation method which enables complete decentralized control of the DC-DC stages based on only local sensor feedback. The AFE stages are controlled by a centralized controller using minimal communication and only the LVDC bus feedback. The DC-DC stages are designed and operated to process power pulsating at the double-line frequency and hence, the energy storage requirement on the MVDC bus is minimized. The proposed converter and control architecture achieves voltage and power balanced operation without dedicated balancing controllers. The rest of the paper is organized as follows. First, the converter topology is presented. Second, the proposed control architecture is introduced. Third, system analysis and control design guidelines are presented. Lastly, simulation and experimental results are presented to validate the proposed method. 


\section{Converter Architecture}\label{sec:sst}

\begin{figure}[htb]
	\makebox[\linewidth][c]{\includegraphics[angle = 0, clip, trim=0cm 0cm 0cm 0cm,  width=0.475\textwidth]{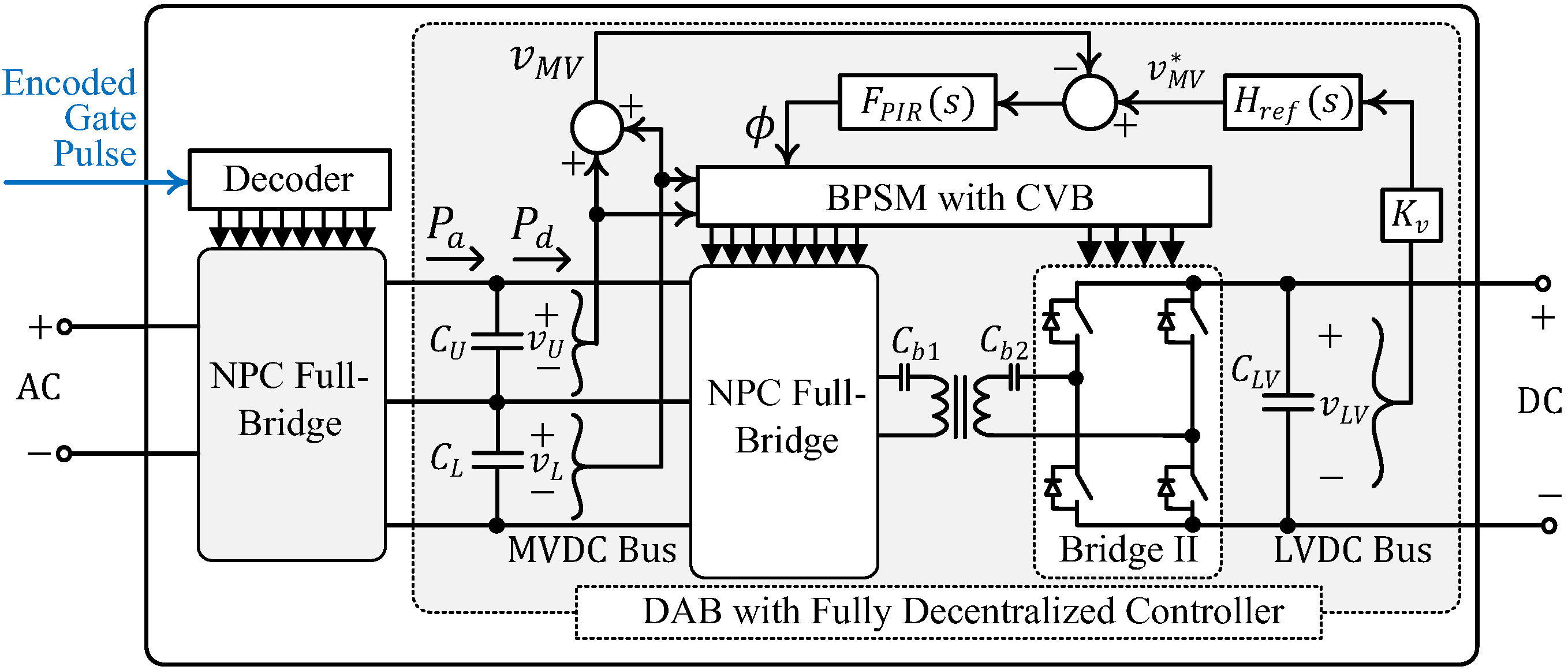}}
	\caption{A single-phase module (SPM) with fully decentralized controller for the DAB stage.}
	\label{fig:spm}
\end{figure}

\begin{figure}[htb]
	\makebox[\linewidth][c]{\includegraphics[angle = 0, clip, trim=0cm 0cm 0cm 0cm,  width=0.475\textwidth]{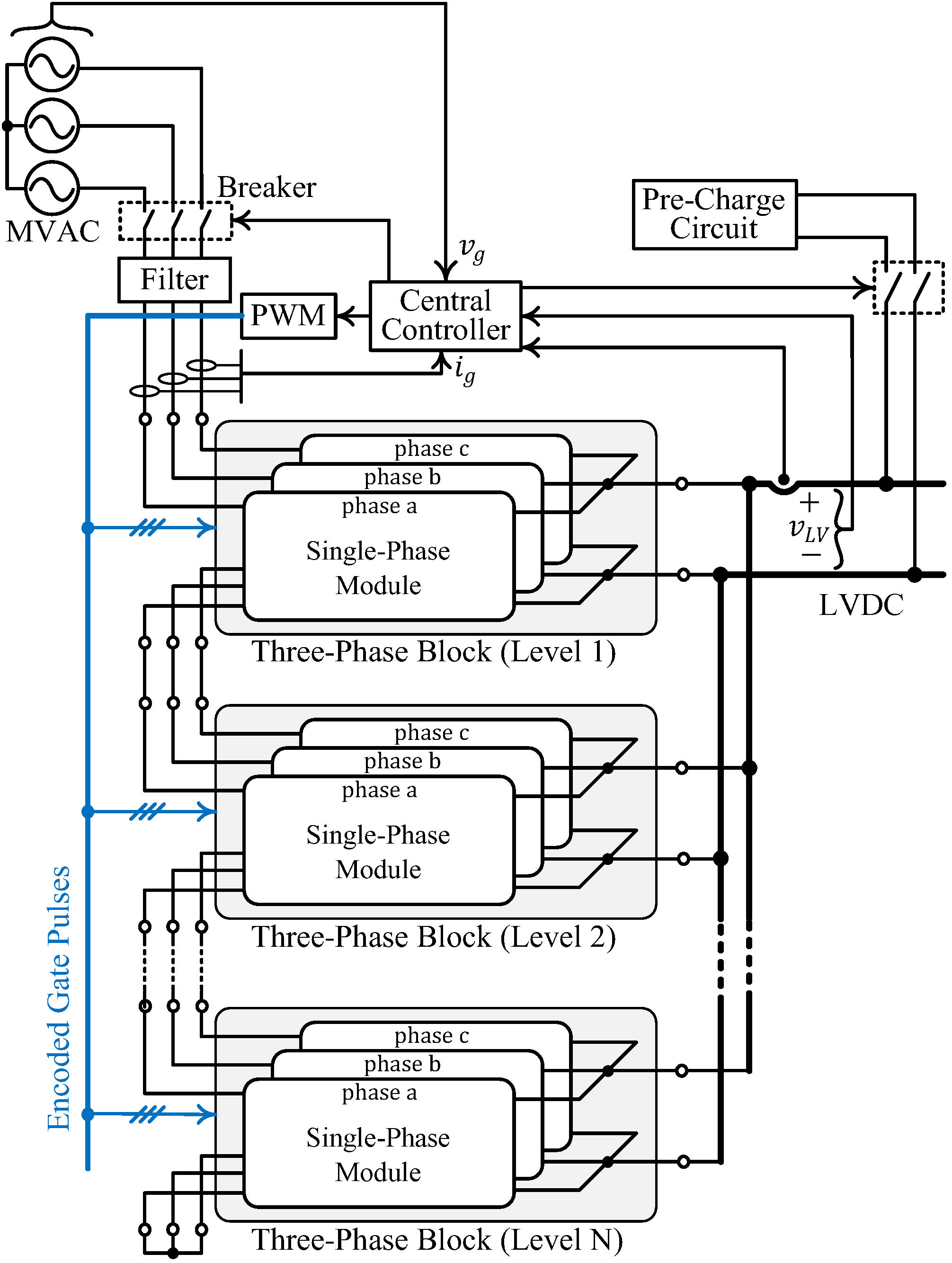}}
	\caption{A three-phase SST configuration using ISOP connected single-phase modules (SPMs); a pre-charge circuit is used for soft start-up.}
	\label{fig:sysDes}
\end{figure}

Galvanically isolated SPMs are used as building blocks for the proposed converter architecture. An SPM using an active-front-end (AFE) stage and an isolated DC-DC stage, i.e., a dual active bridge (DAB) converter, is shown in Fig.~\ref{fig:spm}. A three-level neutral-point diode clamped (NPC) full-bridge topology and a full H-bridge topology are used on the MVDC and LVDC sides of the DAB converter, respectively. DC blocking capacitors $C_{b1}$ and $C_{b2}$ are used to prevent magnetic saturation of the medium frequency transformer (MFT). Another NPC full-bridge is used as the AFE stage. Alternatively, full H-bridge topology can be used for both the MV-side bridge of the DAB and the AFE stage. The DAB stage is designed and controlled as a DC transformer using fully decentralized control, i.e., no communication with other modules or a centralized controller is required. The AFE stage is operated based on encoded gate pulses received via optical fibers. The detail control structure is presented in Section~\ref{sec:Control}. 

The modular converter structure is shown in Fig.~\ref{fig:sysDes}. Each SPM connects to a single-phase AC supply through the AFE and hence subject to a double-line frequency power pulsation. If DC power is drawn through the DC terminal, the MVDC bus requires a large capacitor which serves as an energy buffer to compensate the instantaneous difference between the AC and DC power flows through the AFE and the DAB stages, respectively. To avoid large energy storage element on the MVDC bus, the DAB stage is designed and operated to process AC power, same as the AFE stage; consequently, double line-frequency voltage variation is avoided using minimal capacitor on the MVDC bus. The DC terminals of SPMs with their AC terminals connected to a three-phase supply are tied together in a three-phase block; the AC power pulsations through the SPMs in a three-phase block combines at the DC terminal and constant DC output is obtained. To reach the desired AC voltage level, $N$ three-phase blocks are connected in ISOP configuration (see Fig.~\ref{fig:sysDes}). Common mode (CM) and differential mode (DM) filters are used to meet relevant grid codes. A central controller is responsible for maintaining the LVDC output by dynamically regulating the grid current; the central controller is also responsible soft start-up of the system using a grid-side breaker and a pre-charge circuit. The start-up process is explained in Section~\ref{sec:startUp}.

One of the key challenges for the medium voltage application is to achieve the required basic insulation level (BIL). In typical applications, the LVDC bus should be referenced to the protective earth (PE) ground. In such a configuration, the MFTs in the DAB stages have to be designed to provide the BIL capability. For instance, to connect to a 13.2kV feeder on the AC side, BIL capability of $\approx 90$kV is required. The MFT and the auxiliary systems such as the gate-drivers for the power semiconductor devices and auxiliary power supply for the digital control boards have to be properly designed and grounded to achieve the required BIL capability.    

\section{Control Structure}\label{sec:Control}
In existing control approaches reported in prior art, the grid current is dynamically regulated to maintain the internal MVDC buses, whereas the power flow through the DC-DC stages are controlled to regulate the LVDC bus output. Consequently, module level voltage and power balancing involving a central controller with high speed bidirectional communication to the modules is required. We propose an alternative control structure where the grid current is dynamically controlled through the AFE stages to regulate the LVDC bus voltage. The DAB stages are operated as DC transformers to maintain the MVDC bus voltages keeping constant scaling with respect to the LVDC bus. For the control design, we leverage a well-defined time-scale separation between the MVDC bus and LVDC bus voltage regulation loops, the former achieves at least an order of magnitude faster control response relative to that of the later. The control structures of the DAB and the AFE stages are explained in the following subsections.

\subsection{Control of DAB Stages}
The DAB stage in each SPM uses a fully decentralized controller based on only local sensor feedback to maintain its internal MVDC bus with a control bandwidth of at least an order of magnitude faster than that of the LVDC bus regulation. Hence, for the control design of the DAB converter the LVDC bus voltage $v_{LV}$ is assumed constant in the frequency range of interest and the reference $v^*_{MV}$ for the MVDC bus voltage $v_{LV}$ is dynamically generated as 

\begin{equation}\label{eq:mvdcRef}
  \begin{split}
    v^*_{MV}(s)=K_v H_{ref}(s) v_{LV}(s);\ H_{ref}(s)=\frac{\omega_{ref}}{s+\omega_{ref}},
  \end{split}
\end{equation}

\noindent
where, $K_v$ denotes the constant voltage scaling factor and $\omega_{ref}$ denotes the bandwidth for the reference generation. Bi-directional phase shift modulation (BPSM) is used and a closed-loop voltage compensator generates the phase shift $\phi$ for the BPSM. For an NPC full-bridge on the MVDC side, capacitor voltage balancing (CVB) is required. The detailed design and implementation of the BPSM with CVB can be found in \cite{npcCvbPaper}. Double phase-shift modulation (DPS) is used; however, for simplicity of analysis, $\approx 50\%$ duty ratio is assumed and hence, the power transfer between the two bridges for a phase shift $\phi$ is given as

\begin{equation}\label{eq:Pdab}
  \begin{split}
    P_d = \frac{n v_{LV} v_{MV}}{2\pi f_s L}\{\phi(1 - \phi)\},
  \end{split}
\end{equation}

\noindent
where, $n$ and $L$ denote the turns-ratio and the leakage inductance  of the MFT including any external inductor, respectively; $f_s$ denotes the switching frequency and $\phi>0$ corresponds to MVDC-side bridge output voltage leading that of the LVDC-side bridge. Note that the leakage inductance $L$ is referred to the MVDC side of the MFT. The frequency of the LC series resonance contributed by the leakage inductance and the blocking capacitors is given by

\begin{equation}\label{eq:cblocking}
  \begin{split}
    f_r = \frac{1}{2\pi} \sqrt{\frac{n^2 C_{b1} + C_{b2}}{L C_{b1} C_{b2}}}.
  \end{split}
\end{equation}

\noindent
The blocking capacitors are chosen to set the LC resonance frequency as $(f_{s1}/10)<f_r\ll f_{s1}$, which enables ignoring the LC resonant dynamics in the frequency range of interest such as $< f_s/10$ for the voltage compensator design. The voltage dynamics of the MVDC bus is derived as 

\begin{equation}\label{eq:vmvdcDyn}
  \begin{split}
    \frac{d}{dt}(\frac{1}{2}C_{MV}v^2_{MV}) = P_a - P_d,
  \end{split}
\end{equation}

\noindent
where, $P_a$ denotes the power flow through the AFE bridge (see Fig.~\ref{fig:spm}). Linearizing \eqref{eq:Pdab} and \eqref{eq:vmvdcDyn}, the small signal response of the MVDC bus voltage can be derived as 

\begin{equation}\label{eq:vmvdcSSR}
  \begin{split}
    \frac{\Delta v_{MV}}{\Delta \phi} = -\frac{nV_{LV}}{2\pi f_s L C_{MV} s},
  \end{split}
\end{equation}
\noindent
where, $V_{LV}$ denotes the operating value of $v_{LV}$. 

\begin{figure}[htb]
	\makebox[\linewidth][c]{\includegraphics[angle = 0, clip, trim=0cm 0cm 0cm 0cm,  width=0.375\textwidth]{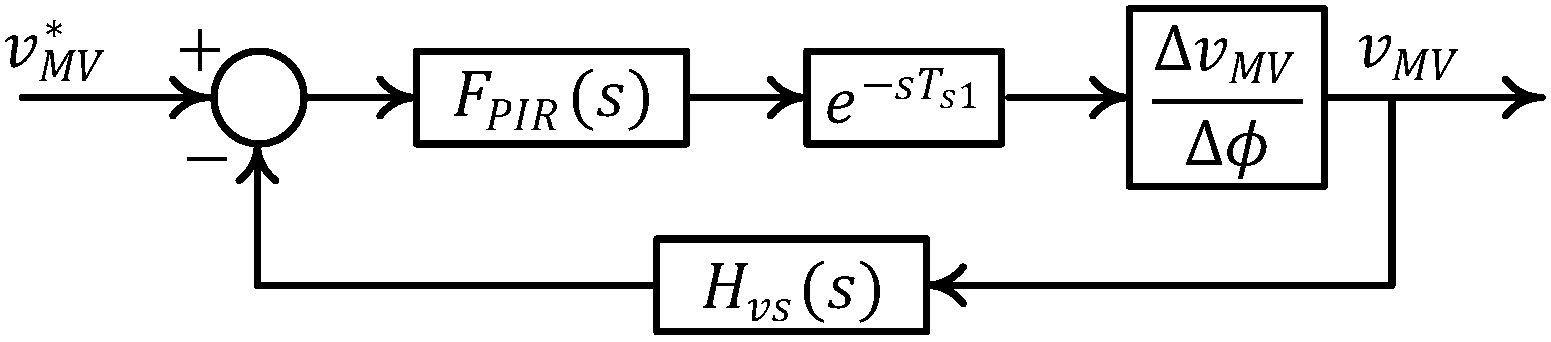}}
	\caption{Control system block-diagram for the DAB stage.}
	\label{fig:dabConSys}
\end{figure}

\noindent
The control system block diagram for the MVDC bus voltage regulation is shown in Fig.~\ref{fig:dabConSys}, where one-sample delay of $T_s$ is considered corresponding to controller implementation. The voltage sensor is modelled as 
\begin{equation}\label{eq:vmvdcSSR}
  \begin{split}
    H_{vs}(s)=\frac{\omega_{vs}}{s+\omega_{vs}}\times e^{-sT_{vs}},
  \end{split}
\end{equation}

\noindent
where $\omega_{vs}$ and $T_{vs}$ denote the bandwidth and transmission time of the voltage sensor. The digital control board for each DAB stage is referenced to the PE ground. Consequently, the required BIL requirement of $\approx 90$kV must be met while sensing the MVDC bus voltage. Therefore, delta-sigma voltage sensors are used and the sensor output is transferred over optical fibers as bit-streams to the digital control board. These optically isolated sensors incur substantial transmission delays on the order of $40\mu $s-$100\mu$s and hence proper modelling of the sensor and signal conditioning is critical for the controller design. The compensator is designed as 

\begin{equation}\label{eq:FPIR}
  \begin{split}
    F_{PIR}(s)=K_{pmv}\left[1+\frac{1}{sT_{imv}} + \frac{(1/T_{rmv})\omega_{bmv}s}{s^2+\omega_{bmv}s+(2\omega_0)^2}\right],
  \end{split}
\end{equation}

\noindent
where, $K_{pmv}$ denotes the proportional gain; $T_{imv}$ is the integral time-constant and $T_{rmv}$ is the time-constant for the resonant compensation at the double line-frequency $2\omega_0$ with a bandwidth of $\omega_{bmv}$. The overall compensated open-loop response is derived as

\begin{equation}\label{eq:Gmvdc}
  \begin{split}
    G_{mvdc}(s)= -\frac{nV_{LV}}{2\pi f_s L C_{MV} s} F_{PIR}(s)H_{vs}(s)e^{-sT_{s1}}.
  \end{split}
\end{equation}

To illustrate the compensator design, we consider the converter system with parameters listed in Table~\ref{TB:sysParam}, whereas the SPM parameters are listed in Table~\ref{TB:spmParam}.   

\begin{table}[htb]
\centering
\caption{Single-Phase Module (SPM) Parameters}
\begin{tabular}{ p{1cm}p{5cm}p{1.5cm}}
\toprule
$V_{MV0}$ & Nominal MVDC bus voltage & $2.15\ \text{kV}$\\
$V^{ac}_{spm}$ & Nominal MVAC voltage & $1.27\ \text{kV}$\\
$V^{dc}_{spm}$ & Nominal LVDC voltage & $750\ \text{V}$\\
$P_{spm}$ & Rated real power & $55.6\ \text{kW}$\\
$Q_{spm}$ & Rated reactive power & $25\ \text{kVAR}$\\
$f_{s1}$ & DAB switching/sampling frequency & $20\ \text{kHz}$\\
$f_{s2}$ & AFE bridge switching frequency & $5\ \text{kHz}$\\
$C_{MV}$ & MVDC bus capacitor & $268\ \mu\text{F}$\\
$L$ & Leakage inductor of DAB MFT & $137\ \mu\text{H}$\\
$n$ & MFT turns-ratio & $3$\\
$C_{b1}$ & DC blocking capacitor (MVDC side) & $6.8\ \mu\text{F}$\\
$C_{b2}$ & DC blocking capacitor (LVDC side) & $150\ \mu\text{F}$\\
\bottomrule
\end{tabular}
\label{TB:spmParam}
\end{table}

\renewcommand{\arraystretch}{1.2}
\begin{table}[htb]
\centering
\caption{System Level Parameters}
\begin{tabular}{ p{1cm}p{5cm}p{1.5cm}}
\hline
\toprule
$S_{rated}$ & Rated power & $1.1\ \text{MVA}$\\
$P_{rated}$ & Rated real power & $1\ \text{MW}$\\
$Q_{rated}$ & Rated reactive power & $450\ \text{kVAR}$\\
$V_{g0}$ & Nominal (L-L RMS) grid voltage & $13.2\ \text{kV}$\\
$\omega_0$ & Nominal frequency & $2\pi(60)\ \text{rad/s}$\\
$V^*_{LV}$ & Nominal LVDC bus voltage & $750\ \text{V}$\\
$N$ & Number of three-phase blocks & $6$\\
$f_{c}$ & Control frequency for LVDC bus regulation & $10\ \text{kHz}$\\
\bottomrule
\end{tabular}
\label{TB:sysParam}
\end{table}

\begin{figure}[htb]
	\makebox[\linewidth][c]{\includegraphics[angle = 0, clip, trim=0cm 0cm 0cm 0cm,  width=0.415\textwidth]{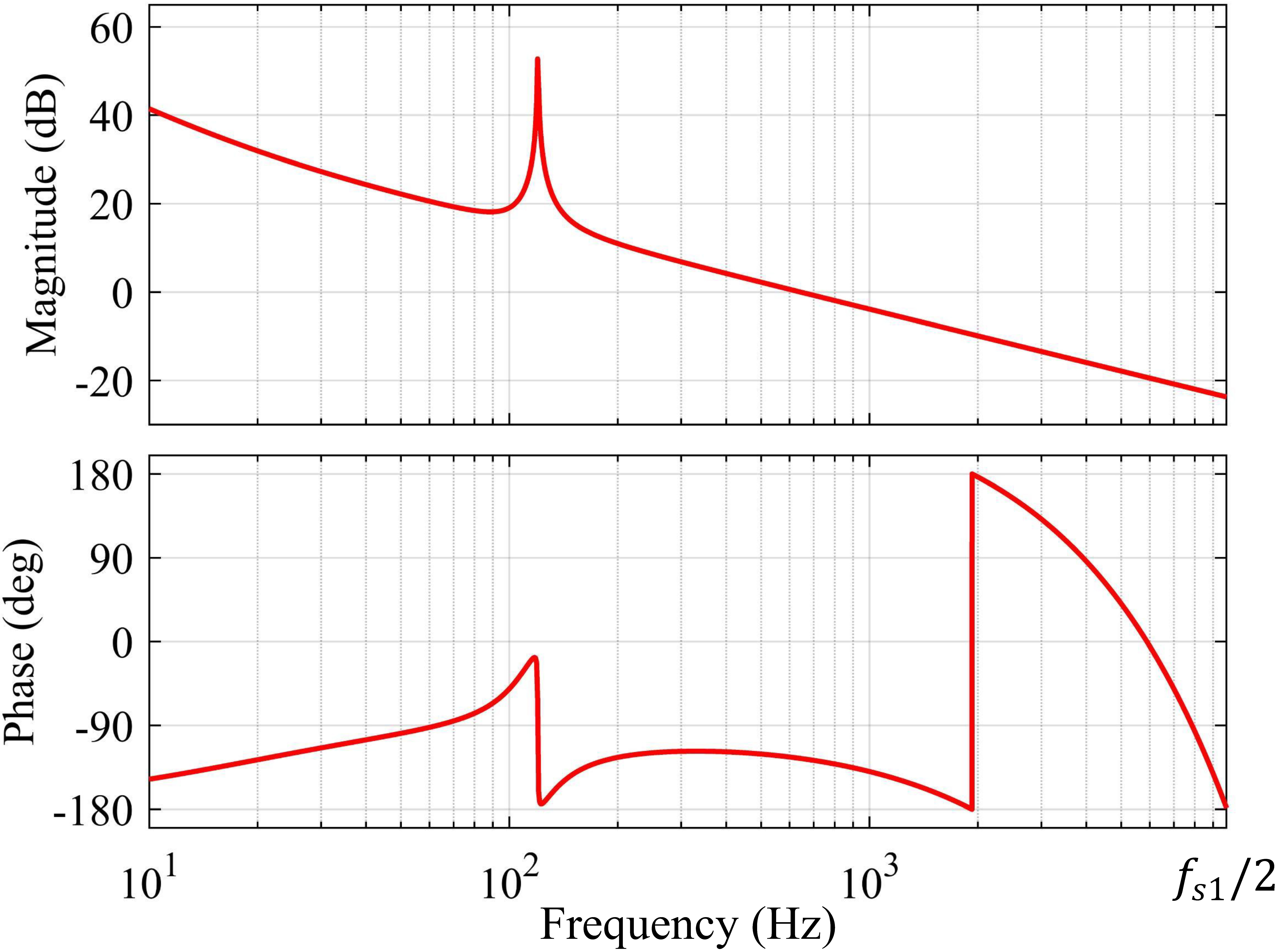}}
	\caption{Compensated open-loop response $G_{mvdc}(s)$ of the MVDC bus voltage regulation.}
	\label{fig:GOLmvdc}
\end{figure}

\noindent
Fig.~\ref{fig:GOLmvdc} shows the compensated open-loop response of the MVDC bus voltage regulation. The delta-sigma voltage sensor has a transmission time of $T_{vs}\approx 77 \mu$s and a bandwidth of $\omega_{vs}\approx 2\pi(100)$ krad/s. The compensator gains are selected as $K_{pmv}=0.0082$ rad/V, $T_{imv}=0.01$ s, $T_{rmv}=0.01$ s, and $\omega_{bmv}=\pi$ rad. Consequently, a control bandwidth of $f_{cmvdc}\approx 643$ Hz is obtained with a phase margin of $55^o$ and gain margin of $10$ dB. The bandwidth of the reference generation filter, given by \eqref{eq:mvdcRef}, is set as $\omega_{ref}/(2\pi)=130\text{ Hz}\approx f_{cmvdc}/5$.

\subsection{LVDC Bus Voltage Regulation}
\begin{figure}[htb]
	\makebox[\linewidth][c]{\includegraphics[angle = 0, clip, trim=0cm 0cm 0cm 0cm,  width=0.4\textwidth]{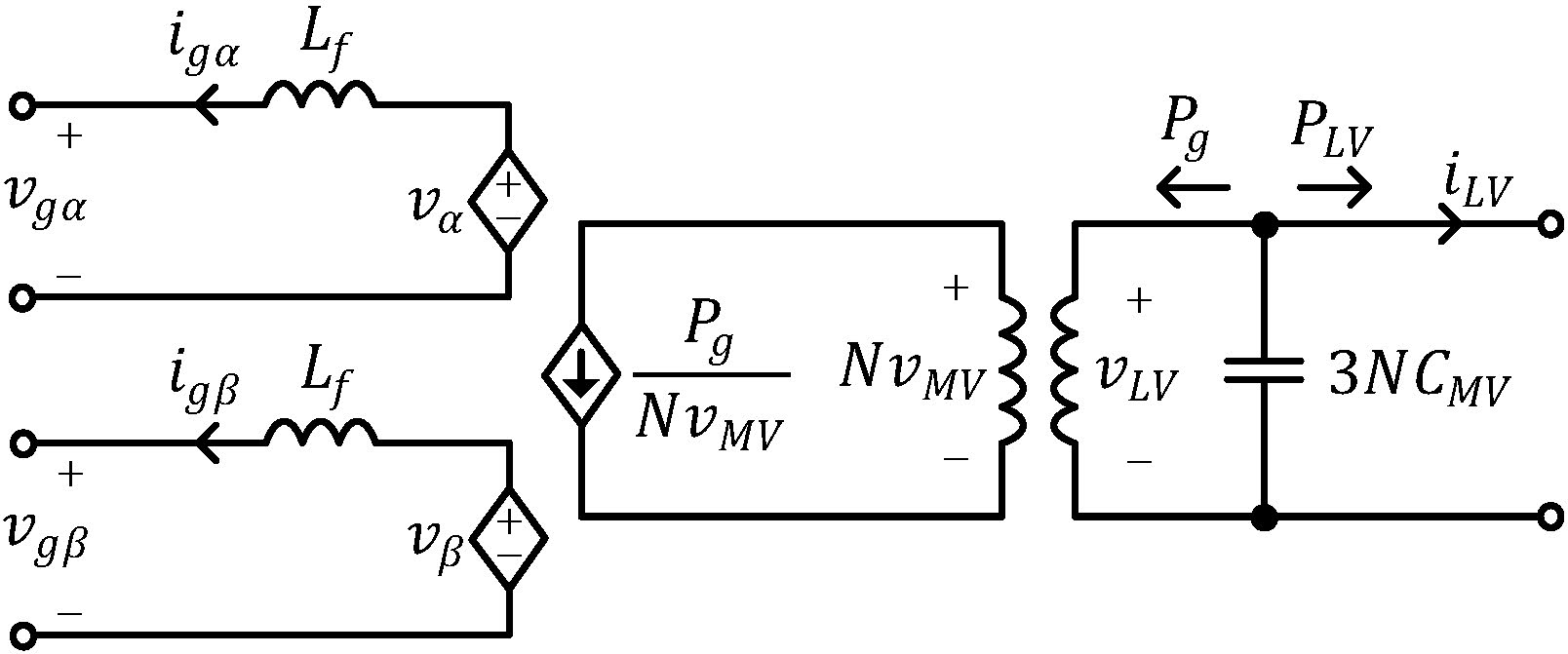}}
	\caption{Equivalent circuit for analysis and design of LVDC bus voltage regulator.}
	\label{fig:lvdvBusConSys}
\end{figure}

In the frequency range of interest, such as tens of Hz, for LVDC bus voltage regulation, the DAB stages effectively behave as DC transformers and provide isolated DC buses for the cascaded H-bridges in the AFE stage. Multi-level modulation or interleaved PWM modulation can be used for the AFE stages. For a device switching frequency of $5$ kHz, multilevel modulation of the $N=6$ AFE bridges in each phase enables to meet relevant grid-codes \cite{std1547} using a simple L-filter as the DM filter. For the analysis and control design, the CM filter is ignored. The equivalent circuit for LVDC bus regulation is shown in Fig.~\ref{fig:lvdvBusConSys}. The system is shown in stationary $\alpha \beta$ frame and the real power injected into the grid is given as $p_g=\frac{3}{2}(v_\alpha i_\alpha + v_\beta i_\beta)$. Note that the ISOP configuration and the multi-level/interleaved modulation facilitate an equivalent MVDC bus voltage of $Nv_{MV}$; therefore, the converter can be effective treated as a three-phase active rectifier followed by an ideal DC transformer supplying DC power to the LVDC bus.

\begin{figure}[htb]
	\makebox[\linewidth][c]{\includegraphics[angle = 0, clip, trim=0cm 0cm 0cm 0cm,  width=0.45\textwidth]{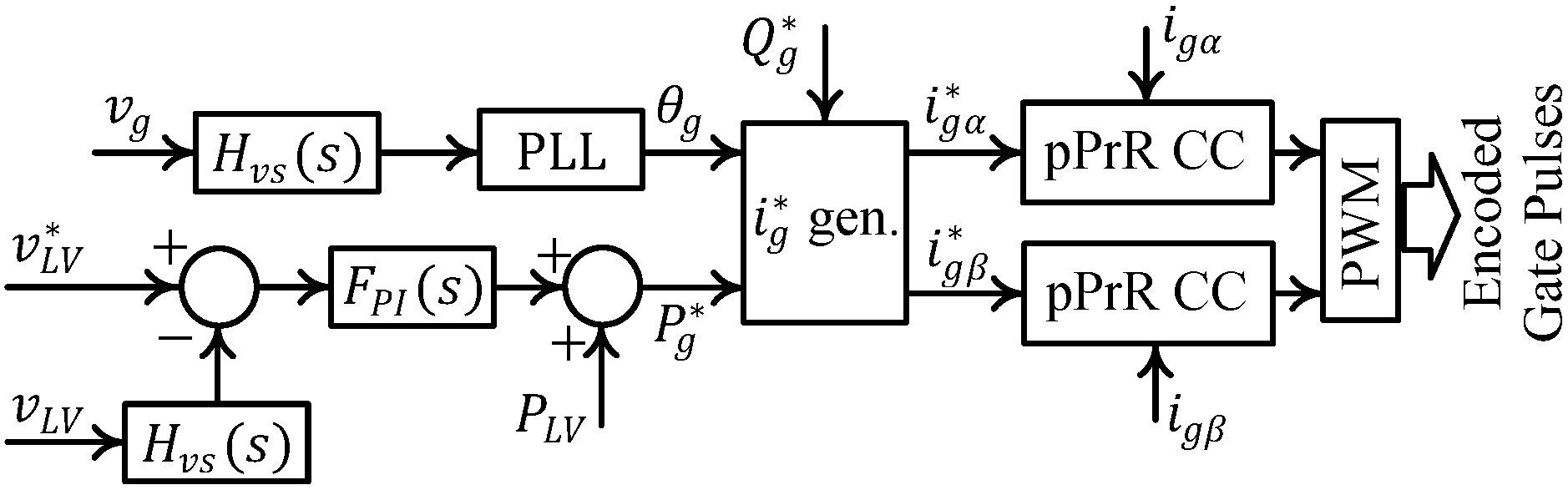}}
	\caption{Control system for LVDC bus voltage regulation.}
	\label{fig:afeConSys}
\end{figure}

\noindent
The LVDC bus voltage dynamics is given as

\begin{equation}\label{eq:lvdcDyn}
  \begin{split}
    \Delta v_{LV}(s)= \frac{1}{3 N C_{MV}V_{MV}}\times [-\Delta P_g(s) - \Delta P_{LV}(s)] ,
  \end{split}
\end{equation}

\noindent
where, $P_{LV}=i_{LV}v_{LV}$ denotes the power drawn from the LVDC bus. A proportional integral (PI) compensator $F_{PI}(s)$ is used to generate the reference $P^*_g$; feedforward of $P_{LV}$ is added using sensor measurement of $i_{LV}$. The control system is shown in Fig.~\ref{fig:afeConSys}. A phase-locked-loop (PLL) running on the grid voltage $v_g$ detects the grid phase $\theta_g$ and the grid-current reference $i^*_\alpha$ and $i^*_\beta$ are generated using $\theta_g$, $P^*_g$, and the reactive power reference $Q^*_g$. Passivity based predictive resonant (pPrR) current controller (CC) is used in the stationary reference frame; the detail design guidelines for the pPrR CC can be found in \cite{pPrR}.     

\noindent
The AFE bridges are used as two-level full H-bridges; the NPC configuration is used only to reach 2.15kV DC bus utilizing lower voltage rated power devices. The PWM state for each AFE bridge corresponding to $\{-v_{MV}, 0, v_{MV}\}$ is encoded into 2-bit digital data and transferred to the respective SPM over optical fibers. The local controller at the SPM decodes the gating information and generates the full gating signals for the power devices. The LVDC bus voltage regulator and the pPrR CC at the central controller are updated at a sampling rate of $f_c=10$kHz. The pPrR CC is designed for a control bandwidth of $\approx 400$ Hz.  

The LVDC bus voltage regulation is performed by a central controller which only uses the sensor feedback of $v_{LV}$ and $i_{LV}$. Unlike existing methods \cite{sstDrB}, the central controller does not require the measurement of all MVDC buses. As the number of SPMs increase to reach higher grid voltage and power capacity, the controller structure does not need to change; only fiber optic cables are added to carry the encoded gate pulses to the oncoming SPMs.

\subsection{Soft Start-Up}\label{sec:startUp}
The ISOP configuration of the converter consisting of $3N$ SPMs poses a unique challenge for system start-up. The MVDC buses in each SPM needs to be charged prior to the closing of the grid-side breaker (see Fig.~\ref{fig:sysDes}) to prevent large current inrush while the AFE bridges behave as uncontrolled rectifier constituted by the anti-parallel diodes of the power devices. Similar inrush problem arises if the DAB stages are started without charging the MVDC buses. Most medium voltage drives with an active-front end typically consists of a dedicated start-up/pre-charge circuit which charges the internal MVDC bus prior to initiating the device switching and connecting to the MV grid. A pre-charge circuit is typically rated at a fractional power capacity compared to the actual converter and provides medium voltage excitation from a low voltage house-keeping supply of $208$VAC or $408$ VAC. Such a pre-charge circuit is also critical for component and system level diagnostics and testing when subject to medium voltage excitation during system assembly phase when a medium voltage feeder is not available. However, for the proposed converter and control structure, a pre-charge circuit (see Fig.~\ref{fig:sysDes}) is used which connects to a low voltage AC supply of $480$ VAC and is used to charge the LVDC bus instead of the MVDC buses. The system start-up sequence is described as follows:

\begin{itemize}
    \item The LVDC bus is charged through the pre-charge circuit; at this condition the grid-side breaker is open, the controller and the power device switching are disabled.
    
    \item The power devices in the LVDC side bridge of the DAB converter in each SPM are switched and the switching duty ratio is ramped up gradually; the MVDC side brdige and the AFE bridge are disabled. The MVDC side bridge of the DAB acts as an uncontrolled rectifier and the MVDC bus is charged without high inrush current. 
    \item Once the MVDC buses are charged, the MVDC side bridge switching is enabled and the closed-loop regulation of the MVDC bus voltage is initiated; the AFE stages are kept disabled. 
    
    \item A low-bandwidth communication channel designed for system monitoring conveys a \emph{ready} signal to the central controller. The pre-charge circuit is disconnected and the grid-side breaker is closed.
    
    \item The LVDC bus voltage regulation and the AFE stages are activated; the system enters nominal operation.
\end{itemize}

Note that a slow/very low-bandwidth communication system is required for the system monitoring and diagnostics which is also used for coordinating the start-up sequence.

\section{Simulation and Experimental Results}
All simulations are performed for the system listed in Table~\ref{TB:spmParam} and Table~\ref{TB:sysParam} in PLECS simulation platform; detailed switching model simulation is used. Voltage and current sensor dynamics are modelled based on the respective hardware sensors. To emulate component tolerances commensurate to the physical system and to illustrate the inherent voltage and power balancing capability of the proposed control architecture, the leakage inductances of the MFTs in the DAB transformers are varied as $L=\{0.91, 0.92,..,1.0,...,1.08\}\times L_0$ for the $18$ SPMs, where $L_0=137\ \mu$H. Similarly, the MVDC bus capacitors are varied as $C_{MV}=\{0.91, 0.92,..,1.0,...,1.08\}\times C_{MV0}$ among the SPMs, where $C_{MV0}=268\ \mu$F. For both $L$ and $C_{MV}$, component values vary between $1\%$ to $17\%$ among different SPMs. 

\begin{figure}[t]
	\makebox[\linewidth][c]{\includegraphics[angle = 0, clip, trim=0cm 0cm 0cm 0cm,  width=0.5\textwidth]{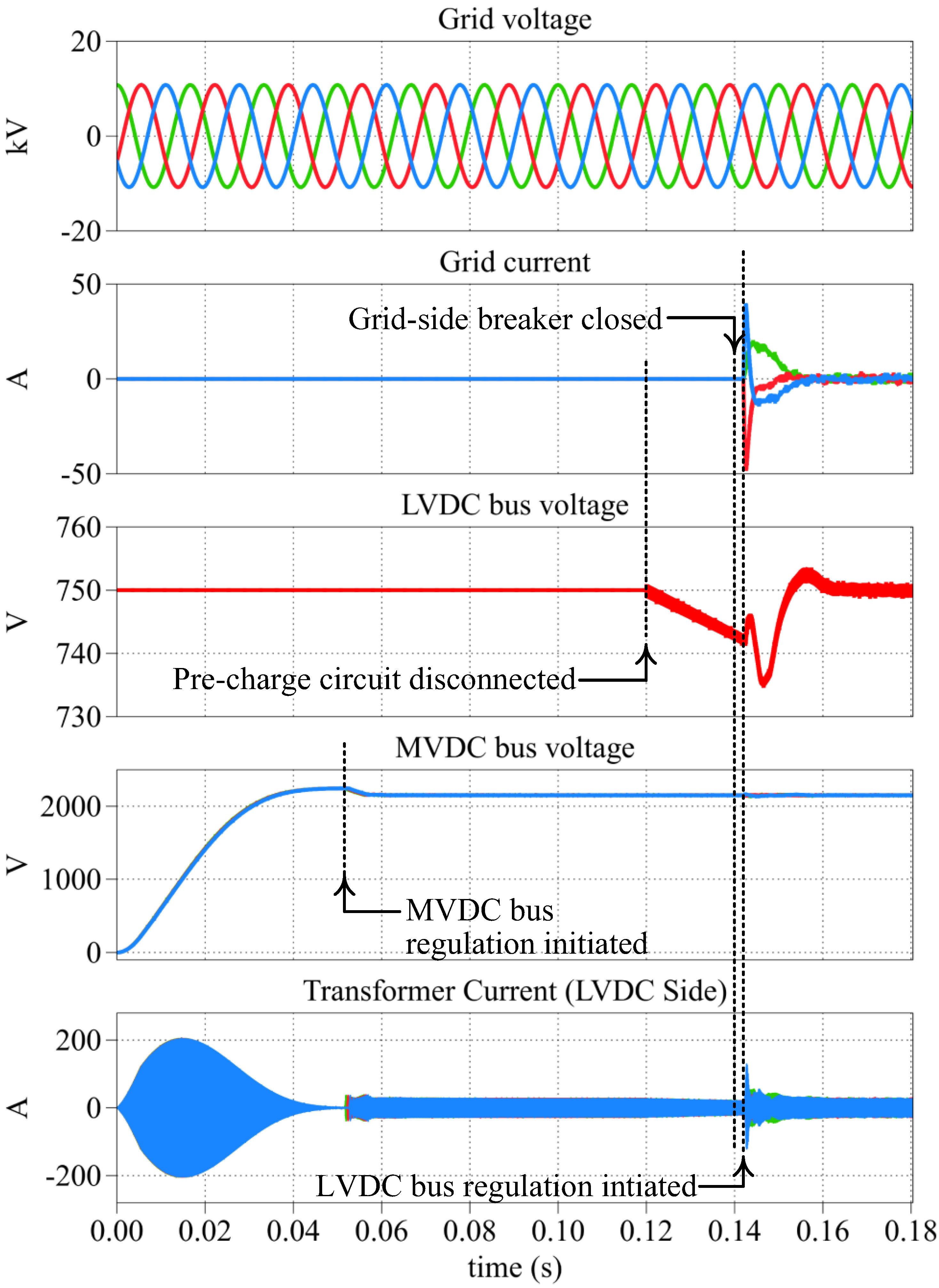}}
	\caption{Soft start-up of the system.}
	\label{fig:simSU}
\end{figure}

\begin{figure}[t]
	\makebox[\linewidth][c]{\includegraphics[angle = 0, clip, trim=0cm 0cm 0cm 0cm,  width=0.5\textwidth]{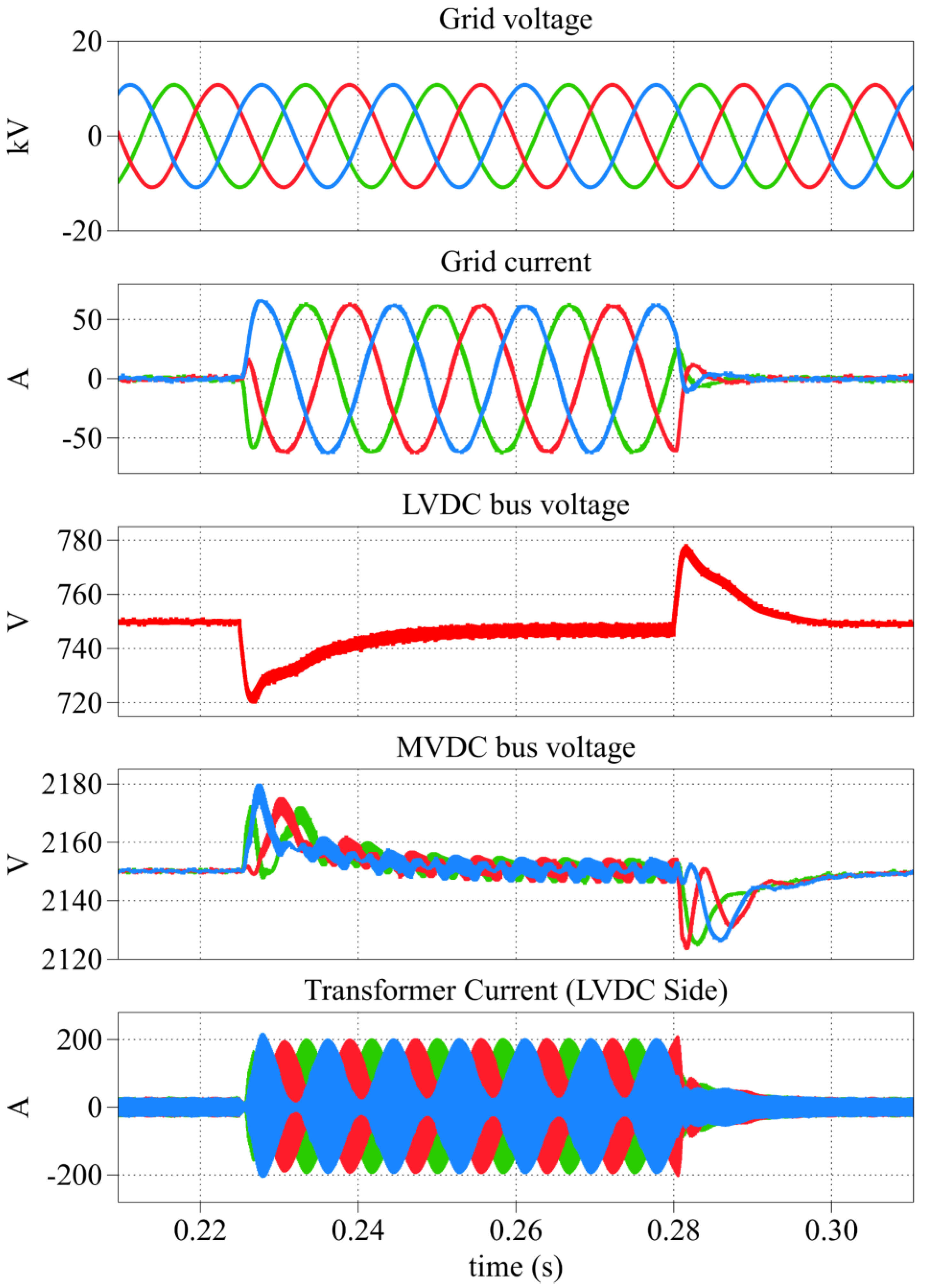}}
	\caption{System response to no-load to full-load and full-load to no-load step change at the LVDC bus.}
	\label{fig:simLT}
\end{figure}

The start-up sequence of the system is shown in Fig.~\ref{fig:simSU}; MVDC bus voltages and LVDC side transformer currents of a three-phase block are shown along with the grid voltage, grid current, and the LVDC bus voltage. Prior to initiating the MVDC bus voltage regulation, the MVDC bus is slowly charged. Once the MVDC bus voltage regulation stabilizes, the pre-charge circuit is disconnected. During this time, the electrical and magnetic losses of the DAB stages are supplied from the LVDC bus capacitors. To emulate such effects, a resistive load of 700W ($\equiv 1.2\%$ p.u. loss) is connected across the MVDC bus; note that the chosen load emulates a very conservative loss since the efficiency of the DAB stage is expected to be $>99\%$. The LVDC bus drops gradually; next, the grid-side breaker is closed without any transient since the MVDC buses are charged. Subsequently, the LVDC bus voltage regulation along with the switching of the AFE bridges is initiated and the LVDC bus voltage is quickly stabilized at the desired value. Fig.~\ref{fig:simLT} shows the system response when no-load to full-load and full-load to no-load step changes are introduced at the LVDC bus. In both cases the LVDC bus voltage is quickly stabilized at the reference value with minimal transients.

\begin{figure}[htb]
	\makebox[\linewidth][c]{\includegraphics[angle = 0, clip, trim=0cm 0cm 0cm 0cm,  width=0.5\textwidth]{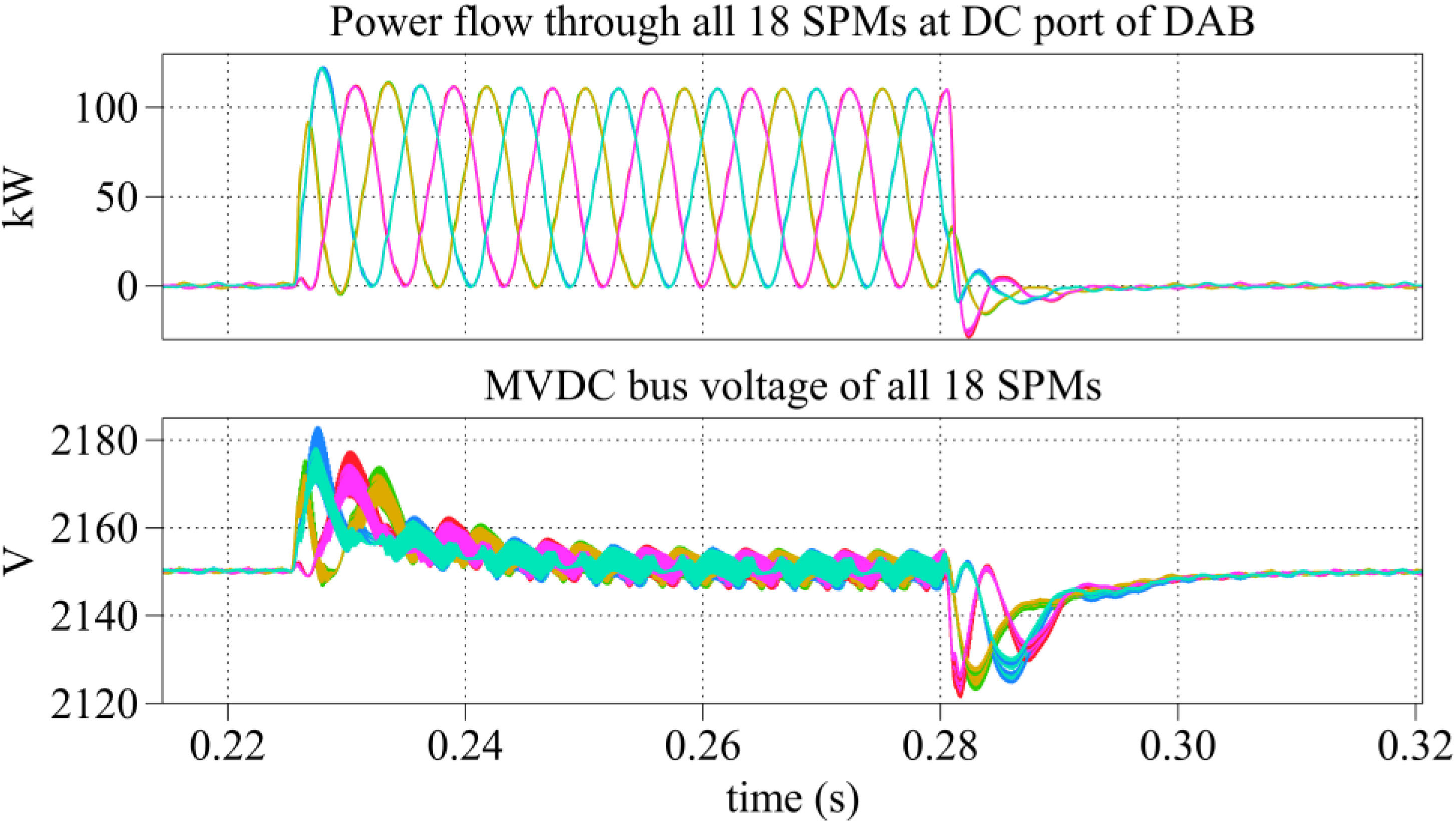}}
	\caption{MVDC bus voltages and power flow through all SPMs are inherently balanced by the controller.}
	\label{fig:simBL}
\end{figure}

The inherent voltage and power balancing among SPMs are shown in Fig.~\ref{fig:simBL}. Despite the variation as high as $17\%$ in the leakage inductance of the MFTs and the MVDC bus capacitor values among different SPMs, the MVDC bus voltage balance and power flow balance is retained even during transients. Note that during transients the MVDC bus voltage and power flow varies among different phases which is expected, but balanced operation is retained among all SPMs corresponding to each phase. 

\begin{figure}[htb]
	\makebox[\linewidth][c]{\includegraphics[angle = 0, clip, trim=0cm 0cm 0cm 0cm,  width=0.5\textwidth]{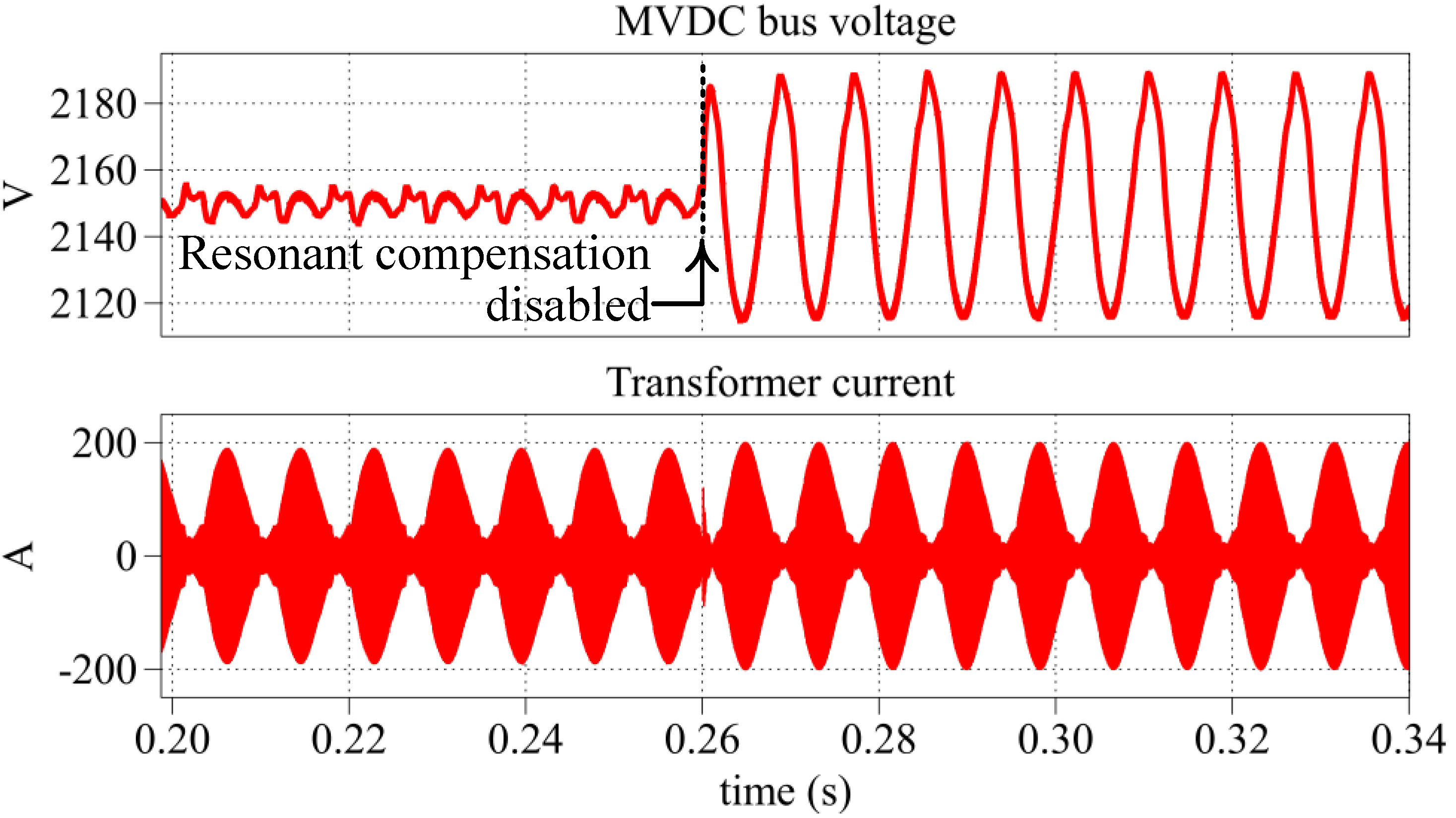}}
	\caption{Disabling the resonant compensation leads to very high ripple in MVDC bus voltage at double-line frequency.}
	\label{fig:simRC}
\end{figure}

The effectiveness of the resonant compensator for the MVDC bus voltage regulation is illustrated in Fig.~\ref{fig:simRC}. The resonant compensator effectively eliminates the double line frequency voltage ripple on the MVDC bus, whereas disabling the resonant compensation leads to $\approx 90$ V ripple. 

\begin{figure}[htb]
	\makebox[\linewidth][c]{\includegraphics[angle = 0, clip, trim=0cm 0cm 0cm 0cm,  width=0.4\textwidth]{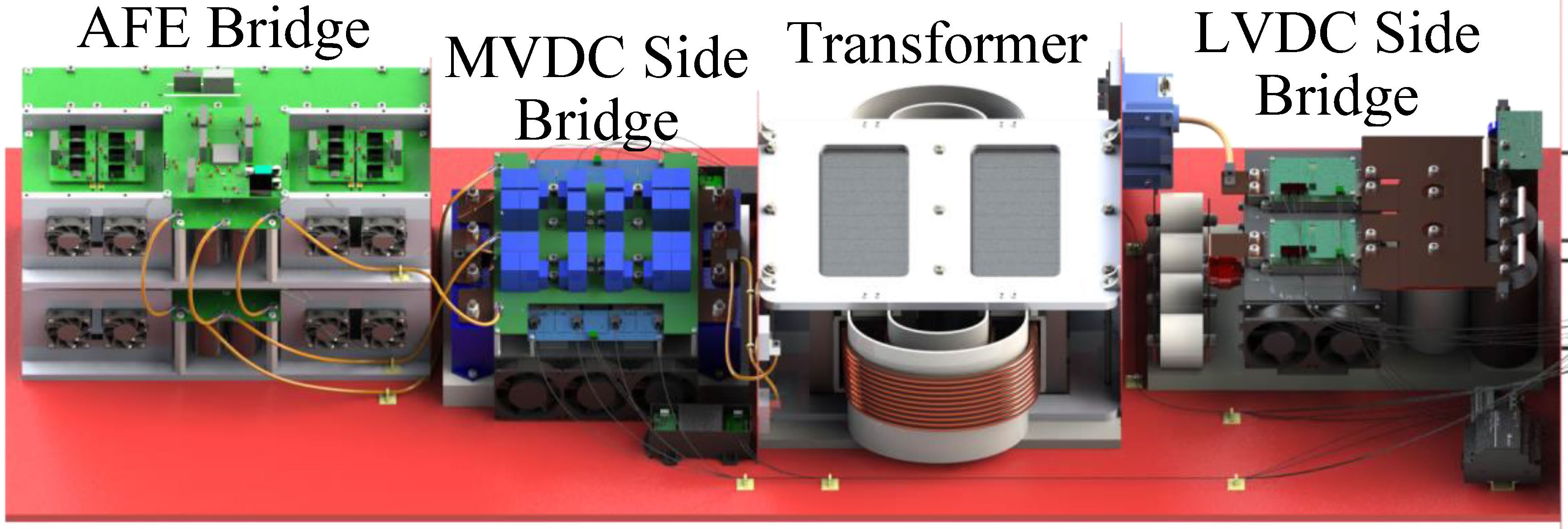}}
	\caption{Hardware prototype assembly of the SPM.}
	\label{fig:spmVA}
\end{figure}

\begin{figure}[htb]
	\makebox[\linewidth][c]{\includegraphics[angle = 0, clip, trim=0cm 0cm 0cm 0cm,  width=0.4\textwidth]{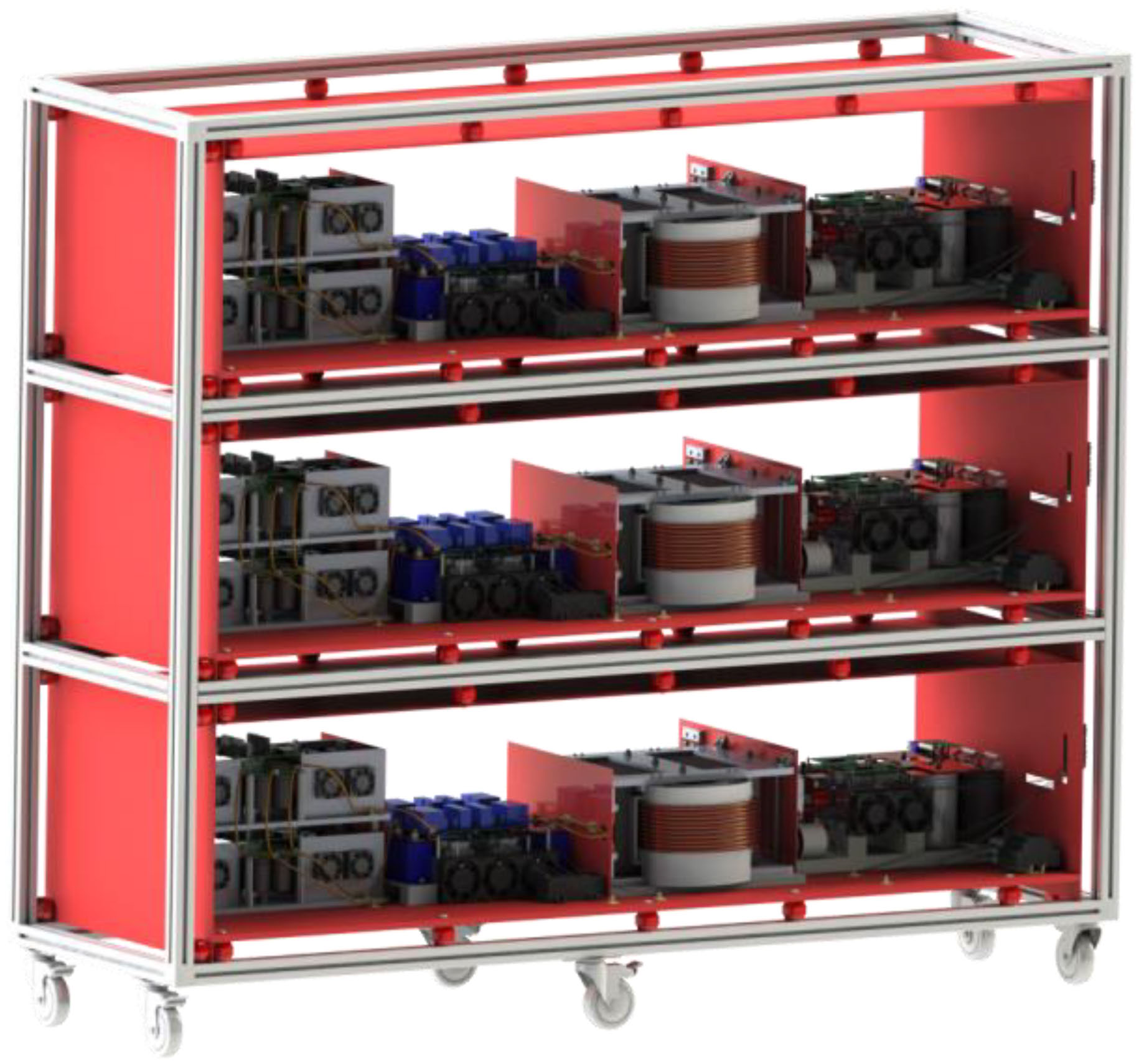}}
	\caption{A three-phase block consisting of three SPMs mounted on a rack.}
	\label{fig:tpbVA}
\end{figure}

\begin{figure}[htb]
	\makebox[\linewidth][c]{\includegraphics[angle = 0, clip, trim=0cm 0cm 0cm 0cm,  width=0.5\textwidth]{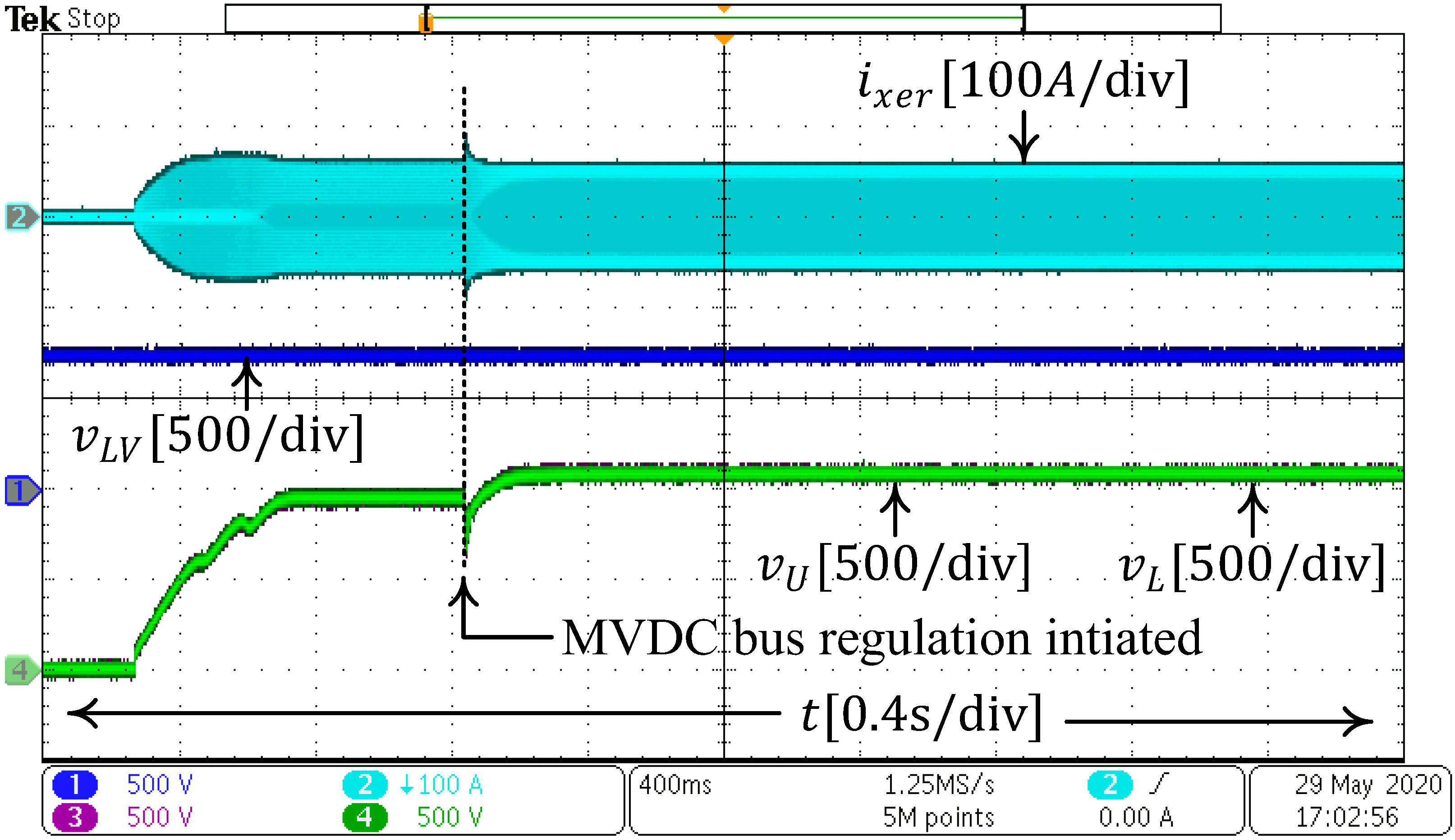}}
	\caption{Soft start-up of the DAB stage.}
	\label{fig:expRes}
\end{figure}

Fig.~\ref{fig:spmVA} shows the prototype assembly of a SPM; a rack-mounted three-phase block consisting of three SPMs is shown in Fig.~\ref{fig:tpbVA}. The soft start-up sequence of the DAB stage is shown in Fig.~\ref{fig:expRes}. The upper and lower capacitor voltages of the MVDC side NPC bridge, the LVDC bus voltage, and the transformer current $i_{xer}$ on the MVDC side are shown. To validate the start-up process and the continuous operation of the DAB, a DC supply is connected to the LVDC bus and a 93kW ($\approx 1.6$p.u.) DC load is connected across the MVDC bus. The MVDC bus voltage is quickly stabilized once the voltage regulator is initiated.  

\section{Conclusion}
The proposed converter and control structure achieves module level voltage and power flow balancing utilizing complete decentralized control of the isolated DC-DC stages. Leveraging proper time-scale separation among different control loops, the DC-DC stages are operated as ideal DC transformers which facilitate isolated DC buses for the AFE stages. LVDC bus voltage regulation is achieved by a central controller using minimal communication. The inherent balancing capability is demonstrated through detailed switching model simulation. The soft start-up is validated through experiments using a full-scale module, whereas the overall system validation is ongoing and will be reported in the final version of the paper. 

\bibliographystyle{IEEEtran}
\bibliography{xfc}

\end{document}